# Decentralized Robust Interval Type-2 Fuzzy Model Predictive Control for Takagi-Sugeno Large-Scale Systems

Mohammad Sarbaz[1], Iman Zamani[1], Mohammad Manthouri[1*], Asier Ibeas[2]

[1]Electrical and Electronic Engineering Department, Shahed University, Tehran, Iran
*Corresponding author. Tel.: +98 21 51212029.
[2]Departament de Telecomunicació i Enginyeria de Sistemes, Escolad'Enginyeria. Universitat Autònoma de Barcelona, Barcelona, Spain
*E-mail addresses: mohammad.sarbaz@shahed.ac.ir (M. Sarbaz), zamaniiman@shahed.ac.ir (I. Zamani), mmanthouri@shahed.ac.ir (M. Manthouri), asier.ibeas@uab.cat (A. Ibeas).*

**Abstract**

In this manuscript, decentralized robust interval type-2 fuzzy model predictive control for Takagi-Sugeno large-scale systems is studied. The mentioned large-scale system consists a number of interval type-2 (IT2) fuzzy Takagi-Sugeno (T-S) subsystems. An important matter and necessities that limit the practical application of model predictive control are the online computational cost and burden of the existence frameworks. For model predictive control of T-S fuzzy large-scale systems, the online computational burden is even worse and in some cases, they cannot be solved in an expected time. Especially for severe large-scale systems with disturbances, existing model predictive control of T-S fuzzy large-scale systems usually leads to a very conservative solution. So, researchers have many challenges and difficulties in finding a reasonable solution in a short time. Although, more relaxed results can be achieved by the proposed fuzzy model predictive control approach which adopts T-S large-scale systems with nonlinear subsystems, many restrictions are not considered in these approaches. In this paper, challenges are solved and the MPC is designed for a nonlinear IT2 fuzzy large-scale system with uncertainties and disturbances. Besides, online optimization problem is solved and results are proposed. Consequently, online computational cost of the optimization problem is reduced considerably. At the end, by two practical examples, the effectiveness of the proposed algorithm is illustrated.

## 1. Introduction

Since years ago, the main task in engineering has been control of processes like mechanical engineering, electrical engineering, chemical engineering, or so. Researchers have proposed many algorithms and approaches in dealing with the instability of systems. But since decades ago, systems have been turned to large in scope and dynamic. Consequently, control of the process has become an essential task. Various controllers have been designed to encounter the instability of systems in both industry and academic like adaptive control, fuzzy control, etc. To design an effective and authentic controller, the dynamic of the system should be identified exactly and it is an important step. In large-scale systems, it is almost impossible to identify the dynamic of the system accurately. Hence, the well-known method, fuzzy logic is used. A useful controller for large-scale systems which has a potential capability to control the process is model predictive control (MPC). Due to the fact that model predictive control uses a cost function to compute the input vector, it has been practical and popular for many years.

Todays, almost all real systems, which are investigated in both industry and academic, are large-scale, and have been broad appeal for decades [1] and [2]. Besides, many systems specially control systems have become large in scope and complex in computation [3]. These type of the systems are constructed by a number of independent subsystems which work independently with some known and unknown interactions. Since years ago, lots of researches have been done in large-scale systems and gain many attentions [4] and [5]. Recently, fuzzy systems with **IF-THEN** rules have become more prevalent and most of the nonlinear and complex systems are modeled by fuzzy logic [6] and [7]. One of the most powerful approaches, which can fill the gap between linear and severe nonlinear systems is well-known Takagi-Sugeno fuzzy model. Many investigations have been done based on the T-S fuzzy model [8] and [9]. [10] and [11] propose a type of fuzzy inference system popular Takagi-Sugeno, fuzzy model. In [12], Takagi-Sugeno fuzzy model is selected to represent the dynamic of the unknown nonlinear system. But in aforementioned researches, uncertainties are not involved in membership functions for the type-1 fuzzy set. So the control problem for nonlinear plants subject to uncertainties are not handled openly. If uncertainties of nonlinear plants result in uncertainties grades of membership functions, conditions based on Takagi-Sugeno type- 1 might be directed to the conservativeness. Here, the interval type-2 fuzzy Takagi-Sugeno is introduced to fix uncertainties captured by the interval type-2 membership functions [13] and [14]. In fact, IT2 has many boons in handling the grades of membership uncertainties over type-1. Applicable IT2 has been used widely in control algorithms like [15] and [16], which nonlinear network control system has been modeled by IT2 for synthesizing approach of dynamic output feedback MPC. In [17] the adaptive sliding mode control problem is introduced for the uncertain nonlinear system, which is modeled by IT2 Takagi-Sugeno fuzzy model.

Recently, MPC has attended popularity as a reliable control approach. By properly using the system model to predict the output response, MPC methods allow to choose the optimal control action that minimizes a desired cost function. In many applications, MPC has been recognized as a viable alternative to many other classic schemes based. For such applications, the advantages introduced by MPC are the higher dynamic performance, the possibility of performing a multivariable controller design, the inclusions of constraints on input and output variables, and the possibility of including nonlinearities in both the model and the constraints. Although the need for an accurate model may represent a drawback in some applications, it is remarked how reliable model of systems are usually available for control design [18] and [19]. The gist of the model predictive control is an optimal control sequence, which is computed by minimizing a finite horizon cost function at each sampling time. Although for many years lots of researches have been done without considering the uncertainties and disturbances [20], in real systems both uncertainties and disturbances exist, and this may lead to inconstancy. So the robust model predictive control should be considered in this circumstance, and by using the robust $H_\infty$ strategy, we try to diminish the effect of disturbances and uncertainties [21] and [18]. Many works have been studied in nonlinear robust MPC, e.g., [22] and [23]. In [24] both online and off-line robust fuzzy model predictive control with structured uncertainties and persistent disturbances is investigated for usual systems. In [18], robust fuzzy model predictive control with nonlinear local models is introduced, in which for both nonlinear and linear states of the system, separate controllers are proposed. The integration of photovoltaics into distribution power systems with grid fault ride-through capability is investigated by proposing a robust model predictive control scheme in [25]. After that, many studies have been improved and can be found in [26]. An important avenue to MPC is based on linear matrix inequalities (LMIs) technique, which is proposed in [27] for linear parameter varying (LPV) systems. In [28] new MPC is contrived for polytypic linear parameter varying systems and paradigm used is adopted in gain scheduling.

Obviously it seems that MPC for interval type-2 fuzzy large-scale systems with persistent disturbances has not been studied yet and several problems remind unsolved. Designing the MPC for fuzzy large-scale system with nonlinear and complex dynamic considering the uncertainties and disturbances has turned to a real challenge and almost in all systems it is impossible to reach a non-conservative solution, specifically solving the online optimization problem. Consequently, in this paper, by the proposed method, this challenge is solved and the MPC is designed for a nonlinear fuzzy large-scale system with uncertainties and disturbances. Besides, online optimization problem is solved and results are illustrated. So the contribution of this paper can be summarized as: (1) An optimal control law is obtained at each sampling instant by solving an online optimization problem. (2) Proposing the interval type-2 fuzzy model predictive control for discrete nonlinear large-scale system. (3) More relaxed and robust results can be achieved by the proposed method. (4) Using the interval type-2 fuzzy Takagi-Sugeno model for large-scale system. (5) Using $H_\infty$ performance to encounter the disturbance.

The rest of this paper is constructed as follows, in Section 2, some preliminaries information about robust interval type-2 fuzzy model predictive control of Takagi-Sugeno systems are introduced. In Section 3, robust positive invariance (RPI) and the computation of terminal constraint set for nonlinear model-based fuzzy systems are provided. In Section 4, a numerical example is offered. Finally, the concluding remarks are given in Section 5.

## 2. Preliminaries

In this part, some preliminaries will be given. First, the input-to-state stability is presented. Second, the system description, interval type-2 fuzzy Takagi-Sugeno large scale system, and the model predictive control are proposed. Furthermore, the prediction model, cost function, and robust positive invariance (RPI) set are introduced.

2.1 Useful definitions and lemmas

Input-to-state stability (ISS) [29] is a stability notion broadly used to assess the stability of nonlinear control systems with external inputs. In the following, some practical definitions and lemmas are provided:

*Definition 1[18]:* $R, R_+, Z$, and $Z_+$ illustrate sets of real numbers, positive real numbers, integers, and positive integers, respectively. $Z_{[m,n]}$ is Symbol of set of integers in the interval $[m, n]$ for convenience. $\|x\|$ depicts the norm of vector $x_i \in R^c$. A real-valued scalar function $\kappa: R_+ \to R_+$ is an M-function ($\kappa \in M$), if it is continuous, rigidly increasing and $\kappa(0) = 0$. So we can say, $\kappa \in M_\infty$ if $\partial \in M$ and $\lim_{s \to \infty} \kappa(s) = \infty$. A function $\beta: R_+ \times R_+ \to R_+$ is an ML-function ($\beta \in ML$), if for each fixed $s > 0, \beta(.,s)$ is a M-function, and for each fixed $r > 0, \beta(r,.)$ is rigidly decreasing and $\beta(r,s) \to 0$ as $s \to \infty$.

*Definition 2 (Input-to-state-Stability):* A discrete-time nonlinear system $x(k+1) = f(x(k), d(k))$ where $d$ is symbol of the disturbance vector, is input-to-state stable (ISS) if there exist an ML-function $\beta$ and a M-function $\gamma$ such that for each input $d$, $\|x(k)\| \leq \beta(x(0), k) + \gamma(\| d \|)$, where $x(0)$ is the initial state vector, $d$ is the disturbance sequence $\{d(0), d(1), \ldots, d(k-1)\}$ and $k \in Z_+$.

*Definition 3 (ISS Lyapunov Function [24]):* A continuous positive definite function $V(x(k))$ is called an ISS-Lyapunov function for system $x(k+1) = f(x(k), d(k))$, if there exist $M_\infty$-function $\kappa_1, \kappa_2, \kappa_3$, and M-function $\rho$ such that:

$$\kappa_1(\| x(k) \|) \leq V(x(k)) \leq \kappa_2(\| x(k) \|) \tag{1}$$

$$V(x(k+1)) - V(x(k)) \leq -\kappa_3(\| x(k) \|) + \rho(\| d(k) \|) \tag{2}$$

*Lemma 1[29]:* If system $x(k+1) = f(x(k), d(k))$ acknowledges an ISS-Lyapunov function, then it is ISS.

## 2.2 System description

The interval type-2 fuzzy large-scale system is assumed, which consists of $N$ subsystems. Here, each subsystem $S_i$, can be designated by an interval type-2 fuzzy Takagi-Sugeno model as shown below:

$$S_i^l: \begin{cases} \textbf{IF } z_{i1} \text{ is } \tilde{F}_{i1}^l \text{ and } \ldots \text{ and } z_{ig} \text{ is } \tilde{F}_{ig}^l \\ \textbf{THEN } x_i(k+1) = A_i^l x_i(k) + B_i^l u_i(k) + E_i^l d_i(k) + \sum_{\substack{j=1 \\ i \neq j}}^{N} g_{ij} x_j(k) \end{cases} \tag{3}$$

where $i = 1, 2, \ldots, N; l = 1, 2, \ldots, r_i$, $A_i^l, B_i^l$, and $E_i^l$ are the system matrices and disturbances of rule-$l$ in subsystem $S_i$; $x_i(k) \in R^c$ the state vector, $u_i(k) \in R^n$ the input vector, $d_i(k) \in R^m$ the addition disturbance. $r_i$, $g_{ij}$, and $F_{iq}^l (q = 1,2,\ldots,g)$ introduce the number of the fuzzy rules in subsystem $S_i$, the interconnection between subsystems, and the linguistic interval type-2 fuzzy sets of the rule $l$ according to the function $z_i(k)$, respectively. $z_i(k) = [z_{i1}, z_{i2}, \ldots, z_{ig}]$ are some measurable premise variables for subsystem $S_i$. Using singleton fuzzifier, product fuzzy inference and center-average defuzzifier. The interval type-2 fuzzy Takagi-Sugeno large-scale system (3) is:

$$x_i^+ = \tilde{A}_{i\mu} x_i + \tilde{B}_{i\mu} u_i + E_{i\mu} d_i + \sum_{\substack{j=1 \\ i \neq j}}^{N} g_{ij} x_j, \quad i = 1, 2, \ldots, N \tag{4}$$

$$\begin{aligned} \tilde{A}_{i\mu} &= \sum_{l=1}^{r_i} w_i^l(z_{iq}) A_i^l \quad ; \quad \tilde{B}_{i\mu} = \sum_{l=1}^{r_i} w_i^l(z_{iq}) B_i^l \\ E_{i\mu} &= \sum_{l=1}^{r_i} w_i^l(z_{iq}) E_i^l \quad ; \quad g_{ij} = \sum_{l=1}^{r_i} w_i^l(z_{iq}) g_{ij} \end{aligned} \tag{5}$$

in which $\underline{w}_i^l(z_{iq}(k)) = \prod_{q=1}^{g} \underline{v}_{\tilde{F}_{i\theta}^l}(z_{iq}(k)) \geq 0$ and $\overline{w}_i^l(z_{iq}(k)) = \prod_{q=1}^{g} \overline{v}_{\tilde{F}_{i\theta}^l}(z_{iq}(k)) \geq 0$ illustrate the lower and upper grades of membership functions, respectively. It is obvious that $\underline{v}_{\tilde{F}_{i\theta}^l}(z_{iq}(k)) \in [0,1]$ and $\overline{v}_{\tilde{F}_{i\theta}^l}(z_{iq}(k)) \in [0,1]$ are lower and upper membership functions, respectively. And here, needless to say that, $\overline{v}_{\tilde{F}_{i\theta}^l}(z_{iq}(k)) \geq \underline{v}_{\tilde{F}_{i\theta}^l}(z_{iq}(k))$, therefore, $\overline{w}_i^l(z_{iq}(k)) \geq \underline{w}_i^l(z_{iq}(k))$.

here, $x_i^+$ illustrates the $i$-th system state in the next instant for simplicity intention. The following interval sets are the firing strength of rule:

$W_i^l = [\overline{w}_i^l, \underline{w}_i^l]$

$$w_i^l(z_{iq}) = \overline{w}_i^l(z_{iq}(k))\bar{\rho}_i^l(x(k)) + \underline{w}_i^l(z_{iq}(k))\underline{\rho}_i^l(x(k)) \tag{6}$$

The parameter uncertainties existing in the nonlinear plant can result in uncertainties of the membership functions and determine the lower and upper membership functions. In our investigation, the lower and upper functions are chosen as nonlinear functions related to the state variables.

where $\bar{\rho}_i^l(x(k)) \in [0,1]$ and $\underline{\rho}_i^l(x(k)) \in [0,1]$ are nonlinear functions and satisfy $\bar{\rho}_i^l(x(k)) + \underline{\rho}_i^l(x(k)) = 1$.

*Remark 1:* It is evident that the parameter uncertainties exist in almost all of the nonlinear plants, and they can be resulted in uncertainties in the membership functions and determine the lower and upper membership functions. In some investigation, $\bar{\rho}_i^l(x(k))$ and $\underline{\rho}_i^l(x(k))$ are selected as known and constant parameters [30]. But, in this paper, they, the lower and upper functions, are chosen as nonlinear functions, related to the state variables. Thus, to facilitate the stability analysis and design of the interval type-2, the lower and upper membership functions can be exerted.

The nonlinear fuzzy model predictive control with the schematic is:

$$C_i^l : \begin{cases} \text{IF } z_{il} \text{ is } \tilde{G}_{i1}^l \text{ and } \ldots \text{ and } z_{ig} \text{ is } \tilde{G}_{ig}^l \\ \text{THEN } u_i(k) = k_i^l x_i(k) \end{cases} \tag{7}$$

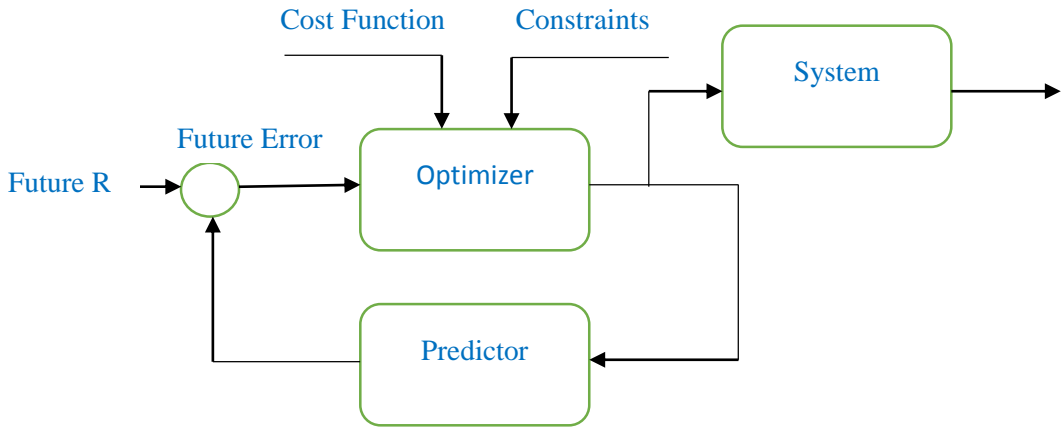

Fig. 1 schematic diagram of the control

where $i = 1,2,\ldots,N; l = 1,2,\ldots,r_i$. For convenience, we use the same weight notation $w_i^l(z_{iq})$ as in subsystem $S_i$. Analogous to (7), the final output of the controller for the corresponding subsystem $S_i$ is:

$$u_i(k) = \sum_{l=1}^{r_i} h_i^l(z_{iq}) k_i^l x_i(k) \tag{8}$$

in which $\underline{h}_i^l(z_{iq}(k)) = \prod_{q=1}^g \underline{\sigma}_{\tilde{G}_{i\theta}^l}(z_{iq}(k)) \geq 0$ and $\overline{h}_i^l(z_{iq}(k)) = \prod_{q=1}^g \overline{\sigma}_{\tilde{G}_{i\theta}^l}(z_{iq}(k)) \geq 0$ illustrate the lower and upper grades of membership functions, respectively. It is obvious that $\underline{\sigma}_{\tilde{G}_{i\theta}^l}(z_{iq}(k)) \in [0,1]$ and $\overline{\sigma}_{\tilde{G}_{i\theta}^l}(z_{iq}(k)) \in [0,1]$ are lower and upper membership functions, respectively. And here, it is needless to say that, $\overline{\sigma}_{\tilde{G}_{i\theta}^l}(z_{iq}(k)) \geq \underline{\sigma}_{\tilde{G}_{i\theta}^l}(z_{iq}(k))$, therefore, $\overline{h}_i^l(z_{iq}(k)) \geq \underline{h}_i^l(z_{iq}(k))$. The following interval sets are the firing strength of rule:

$$h_i^l(z_{iq}) = \frac{\bar{\mu}_i^l(z_{iq}(k))\overline{h}_i^l(x(k)) + \underline{\mu}_i^l(z_{iq}(k))\underline{h}_i^l(x(k))}{\sum_{l=1}^{r_i}\bar{\mu}_i^l(z_{iq}(k))\overline{h}_i^l(x(k)) + \underline{\mu}_i^l(z_{iq}(k))\underline{h}_i^l(x(k))}, \quad H_i^l = [\overline{h}_i^l, \underline{h}_i^l] \tag{9}$$

where $\bar{\mu}_i^l(x(k)) \in [0,1]$ and $\underline{\mu}_i^l(x(k)) \in [0,1]$ are nonlinear functions and satisfy $\bar{\mu}_i^l(x(k)) + \underline{\mu}_i^l(x(k)) = 1$.

*Remark 2:* it is clear that to diminish the conservativeness in designing the controller, it is better to choose the rational values of the weighting coefficient [31]. So, $\bar{\mu}_i^l(x(k))$ and $\underline{\mu}_i^l(x(k))$ are chosen as the nonlinear functions related to the state variables rather than constant ones to diminish the conservativeness.

Here, combining (4) and (8), the closed-loop interval type-2 fuzzy subsystem will be:

$$x_i(k+1) = \sum_{l=1}^{r_i}\sum_{m=1}^{r_i} w_i^l(z_{iq})h_i^m(z_{iq})\left([\tilde{A}_i^l + \tilde{B}_i^l k_i^m]x_i(k) + E_i^l d_i(k)\right) + \sum_{l=1}^{r_i}\sum_{\substack{j=1\\i\neq j}}^{N} w_i^l(z_{iq})g_{ij}x_j(k) \tag{10}$$

### 2.3 Model predictive control

In this part, the prediction model of the system and a specified finite horizon cost function are introduced.

$$x_i(k+m+1|k) = \tilde{A}_{i\mu}x_i(k+m|k) + \tilde{B}_{i\mu}u_i(k+m|k) + E_{i\mu}d_i(k+m|k) + \sum_{\substack{j=1\\i\neq j}}^{N} g_{ij}x_j(k+m|k) \tag{11}$$

and finite horizon cost function which must be minimized is:

$$J(k) = \sum_{i=1}^{N} j_i(k) = \sum_{i=1}^{N}\left(\psi_i(K) + \sum_{m=0}^{T-1}\psi_i(k+m|k) + V_{in}(x_i(k+m|k))\right) \tag{12}$$

where $\psi_i(k+m|k)$ is the stage cost at the predicted time instant. $V_{in}(x_i(k+T|k))$ is called terminal cost, with $V_{in}(\cdot)$ being a positive definite function [32], and the stage cost is selected as:

$$\psi(k) = \sum_{i=1}^{N}\psi_i(k) =$$

$$= \sum_{i=1}^{N}\left(x_i^T(k+m|k)Qx_i(k+m|k) + u_i^T(k+m|k)Ru_i(k+m|k) - \tau_i d_i^T(k+m|k)d_i(k+m|k)\right) \tag{13}$$

where $R$ and $Q$ are fixed real matrices By choosing rational values of the weighs, we will be able to reduce the conservativeness. So, different values of weights are applied iteratively to reach the best result., and $\tau_i$ is a positive scalar. As it is evident, the cost function entails the disturbance, and obviously, the cost function is impressed by the well-known $H_\infty$ control [33]. Thus, it is not suitable to optimize the cost function directly while the disturbance is involved in. So in that case, a min-max approach is chosen that minimizing the worst-case cost function [34]. As well, at the end of the prediction, it needs that the states enter a terminal constraint set to be asymptotic stability. $x_i(k+T|k) \in \Omega_{iw}$ is the terminal constraint set. The online optimization problem can be illustrated as:

$$\min_{u_i(k+m|k)} \max_{d_i(k+m|k)} J_i(k), \tag{14}$$
$$s.t. \quad u_i(k+m|k) \in U_i$$
$$d_i(k+m|k) \in D_i$$
$$x_i(k+m|k) \in \Omega_{iw}$$

and $d_i \in D_i := \{d_i | d_i^T d_i \leq \eta_i^2\}$, $u_i \in U_i := \{u_i | |u_{is}| \leq u_{is.max}\}$ should be satisfied. where $\eta_i^2$ is a positive scalar, $u_{is}$ is the $s$-th element of the inputs, $s \in Z_{[1,m]}$.

## 3. Main results

### 3.1 Robust Positively Invariant Set for the Interval Type-2 Fuzzy T-S Large-Scale Systems

In this part, the robust positively invariant and constraint set are defined. The trajectories of system are bounded robustly and this feature is guaranteed by the controller, as mentioned previously. The RPI set is shown by $\Omega_{iw}$, and

$u_i(k)$ is the corresponding control law. The RPI set property is provided as, $\forall x_i \in \Omega_{iw}, x_i^+ \in \Omega_{iw}$ for all permissible uncertainties and disturbances. $\Omega_{iw}$ is defined as follow for the interval type-2 fuzzy T-S large-scale system (3),

$$\Omega_{iw} := \left\{ x_i \,\middle|\, \sum_{l=1}^{r_i} w_i^l(z_{iq}) x_i^T P_{i\mu} x_i \leq \xi_i \right\} \tag{15}$$

where $P_{i\mu} = \sum_{l=1}^{r_i} w_i^l(z_{iq}) P_i$, and $\xi_i$ are positive scalar and $P_i$ is a positive constant matrix. The controller for IT2 fuzzy Takagi-Sugeno large-scale system is opted as $u_i(k) = \sum_{l=1}^{r_i} h_i^l(z_{iq}) h_i^l x_i(k)$.

*Lemma 2 [24]:* The set $\Omega_{iw}$ is an RPI set if there exists a positive scalar $\lambda_i, (0 < \lambda_i < 1)$, such that,

$$\sum_{i=1}^{N} \left\{ \left( \frac{1}{\xi_i} x_i^{+T} P_{i\mu}^+ x_i^+ - \frac{1}{\xi_i} x_i^T P_{i\mu} x_i \right) - \lambda_i \left( \frac{1}{\eta_i^2} d_i^T d_i - \frac{1}{\xi_i} x_i^T P_{i\mu} x_i \right) \right\} \leq 0 \tag{16}$$

with $P_{i\mu}^+ = \sum_{l=1}^{r_i} w_i^l(x_i^+) P_i$, for all $x_i^+ \in \tilde{A}_{i\mu} x_i + \tilde{B}_{i\mu} u_i + E_{i\mu} d_i + \sum_{\substack{j=1 \\ i \neq j}}^{N} g_{ij} x_j$, $u_i \in U_i$, and $d_i \in D_i$.

Based on the *Lemma 2* and the concept of $\Omega_{iw}$, *Theorem 1* is introduced to assure the trajectories of the system staying in the $\Omega_{iw}$ set.

*Remark 3:* Here, by proposing the *Theorem 1*, two LMIs are going to be solved and if they are feasible, two results are achieved. (1) It will be proved that the considered large-scale fuzzy system is stable in the sense of Lyapunov. (2) Controller gains are computed in the restricted bound, optimally.

For convenience, a table o notation is provided:

| notation | definition | notation | definition |
|---|---|---|---|
| $\xi_i$ | Positive Scalar variable | $P_i$ | Positive matrix |
| $\tau_i$ | Positive scalar | $N_i$ | $N_i = \dfrac{\xi_i}{\eta_i^2}$ |
| $\alpha$ | Positive scalar | $\eta_i^2$ | positive scalar |
| $M_i$ | $\xi_i R$ | $R$ | Positive weight |
| $N$ | Number of subsystems | $Q$ | Positive weight |
| $\lambda_i$ | Positive scalar | $X_j$ | $X_j = \xi_i P_j$ |
| $X_i$ | $X_i = \xi_i P_i$ | Ts | Sampling time |

Table 1. abbreviations and notation

*Theorem 1:* Consider the fuzzy system (3), if there exist positive definite matrices $X_i$ $X_j$, and $X_{i\mu}$, and $\lambda_i$ such that the following matrix inequalities are feasible:

$$\begin{bmatrix} E_{i\mu}^T X_i E_{i\mu} - \xi_i \lambda_i N_i & \star & \star & \cdots & \star & \star \\ \widetilde{\Theta}_i^T X_i E_{i\mu} & N\sqrt{a} \sum_{\substack{j=1 \\ i\neq j}}^{N} g_{ij}^T X_j g_{ij} - (-\lambda_i + 1)X_{i\mu} & \star & \cdots & \star & \star \\ g_{ij}^T X_i E_{i\mu} & (1-\sqrt{\alpha})g_{ij}^T X_i \widetilde{\Theta}_i & -(\alpha-1)g_{ij}^T X_i g_{ij} & \cdots & \star & \star \\ \vdots & \vdots & \vdots & \ddots & \vdots & \vdots \\ g_{iN}^T X_i E_{i\mu} & (1-\sqrt{\alpha})g_{iN}^T X_i \widetilde{\Theta}_i & -(\alpha-1)g_{iN}^T X_i g_{ij} & \cdots & -(\alpha-1)g_{iN}^T X_i g_{iN} & \star \\ 0 & X_i\widetilde{\Theta}_i & 0 & \cdots & 0 & -N^{-1}X_i^T \end{bmatrix} \leq 0 \quad (17)$$

$$\begin{bmatrix} Z_i & k_{i\mu}^T \\ k_{i\mu} & 1 \end{bmatrix} \geq 0 \quad Z_{iss} \leq u_{is,max}^2, s \in Z_{[1,m]} \tag{18}$$

then the set $\Omega_{iw} = \{x_i | x_i^T P_{i\mu} x_i \leq \xi_i\}$ is a RPI set for the interval type-2 fuzzy system (3) corresponding to the feedback control law $u_i(k) = \sum_{l=1}^{r_i} h_i^l(z_{iq}) k_i^l x_i(k)$. $P_{i\mu} = \sum_{l=1}^{r_i} w_i^l(z_{iq}) P_i$.

$Z_{iss}$ is the *s-th* diagonal element of matrix $Z_i$, $N$ represents the number of subsystems, $N_i = \frac{\xi_i}{\eta_i^2}$, $\alpha \geq 2$, and $i,j,l,N,\alpha \in R_+$, and $\widetilde{\Theta}_i = A_{i\mu} + B_{i\mu} k_{i\mu}$.

**Proof.** See **Appendix A**

3.2 The terminal constraint set

In this section, it will be proven by an LMI that the terminal constraint set $\Omega_{iw}$ be the RPI set. The sub-cost function is:

$$\psi_i(\cdot) = \sum_{k=0}^{K-1} \{x_i^T(k) Q x_i(k) + u_i^T(k) R u_i(k) - \tau_i d_i^T(k) d(k)\} \tag{19}$$

a positive definite function (terminal cost function) $V_i(x)$ exists such that $\forall x_i \in \Omega_{iw}$,

$$\gamma_3(\| x_i \|) \leq V(x_i) \leq \gamma_4(\| x_i \|) \tag{20}$$

$$\sum_{i=1}^{N} V(x_i^+) - V(x_i) < -\sum_{i=1}^{N} \psi_i(\cdot) \tag{21}$$

where $\varsigma_3$ and $\varsigma_4$ are $H_\infty$ functions, $V(x)$ is given as:

$$V(x) = \sum_{i=1}^{N} V_i(x) = \sum_{i=1}^{N} \sum_{l=1}^{r_i} w_i^l(z_{iq}) x_i^T P_i x_i \tag{22}$$

The main and core of every proposed control algorithm is definition of a positive function to prove the stability and effectiveness of the proposed algorithm. In the following, a LMI is clarified to prove that the $\Omega_{iw}$ is a terminal constraint set corresponding to the terminal cost.

*Remark 4*: In this section, by proposing the *Theorem 2*, the concept of the robust positively invariant and constraint set are achieved through solving a LMI, and it will be ensured that trajectories of the large-scale system are stable robustly.

*Theorem 2*: Consider the interval type-2 fuzzy Takagi-Sugeno system (3), if (17), (18) and the following matrix inequality is feasible,

$$\begin{bmatrix} E_{i\mu}^T X_i E_{i\mu} - \xi_i \tau_i & \star & \star & \cdots & \star & \star & \star \\ \widetilde{\Theta}_i^T X_i E_{i\mu} & \zeta_i & \star & \cdots & \star & \star & \star \\ g_{ij}^T X_i E_{i\mu} & (1-\sqrt{\alpha})g_{ij}^T X_i \widetilde{\Theta}_i & -(a-1)g_{ij}^T X_i g_{ij} & \cdots & \star & \star & \star \\ \vdots & \vdots & \vdots & \ddots & \vdots & \vdots & \vdots \\ g_{iN}^T X_i E_{i\mu} & (1-\sqrt{\alpha})g_{iN}^T X_i \widetilde{\Theta}_i & -(a-1)g_{iN}^T X_i g_{ij} & \cdots & -(a-1)g_{iN}^T X_i g_{iN} & \star & \star \\ 0 & M_i k_{i\mu} & 0 & \cdots & 0 & -M_i & \star \\ 0 & X \widetilde{\Theta}_i & 0 & \cdots & 0 & 0 & -N^{-1} X_i^T \end{bmatrix} < 0 \quad (23)$$

then $\Omega_{iw}$ is a terminal constraint set corresponding to the terminal cost function $V_i(x)$. Where $N$ represents the number of subsystems, $\zeta_i = N\sqrt{a}\sum_{\substack{j=1 \\ i\neq j}}^{N} g_{ij}^T X_j g_{ij} - X_{i\mu} + \xi_i Q$, $\widetilde{\Theta}_i = A_{i\mu} + B_{i\mu} k_{i\mu}$, $M_i = \xi_i R$, $\alpha \geq 2$, $X_i = \xi_i P_i$, $X_{i\mu} = \xi_i P_{i\mu}$, $X_j = \xi_i P_j$, and $i, j, l, N, \alpha \in R_+$.

**Proof.** See **Appendix B**

3.3 Control algorithm

According to the aforementioned results and to accomplish the control design, in this section, the online control algorithm is studied. So, the terminal constraint set $V(x(k))$, see (21), should satisfy the following condition,

$$V(x(k)) = \sum_{i=1}^{N} V_i(x(k)) = \sum_{i=1}^{N} \sum_{l=1}^{r_i} w_i^l(z_{iq}) x_i^T P_i x_i \leq \xi_i \quad (24)$$

the following optimization problem, thus, is discussed to minimize $\xi_i$:

$$\min \xi_i \text{ subject to } V_i(x(k)) \leq \xi_i \quad (25)$$

in addition, a sufficient condition for $V_i(x(k)) \leq \xi_i$ is $x_i^T(k) P_i x_i(k) \leq \xi_i$. which is $\xi_i - x_i^T(k) P_i x_i(k) \geq 0$ and equal to $\xi_i - x_i^T(k) \frac{\xi_i P_i}{\xi_i} x_i(k) \geq 0$.

by defining symmetrical matrix $X_i = \xi_i P_i$, is guaranteed by the following LMIs,

$$\begin{bmatrix} \xi_i & x_i^T(k) \\ x_i(k) & X_i^{-1}\xi_i \end{bmatrix} \geq 0 \quad (26)$$

*Note:* The term $X_i$ is a fix value and can be appear as a negative form in the LMI.

*Remark 5:* The principal part of the model predictive control is optimizing a cost function. In this paper, the considered system is the type of large-scale one, and the system is stabilized based on the minimized valued of $\xi_i$. In the previous *theorems*, the bilinear matrix inequalities have been proposed. Because of the fact that the significant part of this paper is minimizing, and to reduce the computational burden, first try different values of $\lambda_i, X_{i\mu}, X_i$, and $X_{i\mu}$, and then try to obtain the minimum amount of $\xi_i$.

**Algorithm**

**Step 1:** Set different values of $M_i$, $N_i$, $X_i$, $\lambda_i$. And Obtain the system state $x_i(k)$.
**Step 2**: Solve the following optimization problem

$$\min_{k_{i\mu}, \varepsilon_i, \nu_i, Z_i} \xi_i, \quad (27)$$

subject to (17), (18), (23), (26),

and for each subsystem, find the values of $k_{i\mu}$ and $X_{i\mu}$. Move the time instant from $k$ to $k+1$ and go to Step 1.

One of the principal elements of the model predictive control is recursive feasibility. Since the MPC is supposed for discrete-time systems, it is notably essential to confirm that if it is feasible for all times or not. In the following *theorem*, this initial task is examined.

*Theorem 3*: In system (4), if the solution of the optimization problem can be attainable at time 0, the solution will be obtainable at any time. So the recursive feasibility performed.

*Proof:* If (27) is feasible at time $k$, then by the $x_i(k) \in \Omega_{iw}$, and according to the theorem 1, which has been implied that $x_i(k+1) \in \Omega_{iw}$, then it can be concluded that (27) can be solvable at time $k+1$. Besides, the solution which achieved at the time $k$, is feasible at time $k+1$, on the other hand, the optimization problem is feasible at all time.

*Theorem 4*: consider that the optimization problem is feasible at the initial time 0, system (4) is **ISS** due to the disturbance $d$.

*Proof:* Assume the Lyapunov function $V(x_i(k)) = x_i^T(k) P_{i\mu} x_i(k)$, in which $P_{i\mu} = \sum_{l=1}^{r_i} w_i^l(z_{iq}) P_i$, here, $P_i$ is defined based on the (22), then it can be achieved,

$$\varpi_{min}^* \|x_i(k)\|^2 \leq V(k, x_i) \leq \varpi_{max}^* \|x_i(k)\|^2 \tag{28}$$

in which
$\varpi_{max}^* = max\{\varpi_{max}(P_i(k)) | i \in Z_{[1,L]}, k \in R\}$
$\varpi_{min}^* = min\{\varpi_{min}(P_i(k)) | i \in Z_{[1,L]}, k \in R\}$
where $\varpi_{max}(\cdot)$ and $\varpi_{min}(\cdot)$ are, respectively, the maximal and minimal eigenvalues.
in addition, (22) implies that,

$$V_k(x_i(k+1)) - V(x_i(k)) < -\left(x_i^T(k) Q x_i(k) + u_i^T(k) R u_i(k) - \tau_i d_i^T(k) d(k)\right) \tag{29}$$

where $V_k(x_i(k+1)) = x_i^T(k+1) P_{i\mu} x_i(k+1)$. And if

$$V_k(x_i(k+1)) - V(k, x_i) < -x_i^T(k) Q x_i(k) + \tau_i d_i^T(k) d(k) \tag{30}$$

due to the (38) at time $(k+1)$:

$$V_{k+1}(x_i(k+1)) \leq V_k(x_i(k+1)) \tag{31}$$

ultimately, it achieves that

$$V_{k+1}(x_i(k+1)) - V(k, x_i) < -x_i^T(k) Q x_i(k) + \tau_i d_i^T(k) d(k) \tag{32}$$

resorting to definition (3), (27) and (32) conduced to the result that $V(x_i(k))$ is an ISS Lyapunov function. On the other hand, the closed-loop system is ISS due to disturbances. So the proof is completed. ∎

*Remark 6:* In [18] and [24], Robust fuzzy model predictive control for a class of fuzzy systems have been studied. The type of systems which has been studied beforehand was discrete-time systems. But in this paper, the decentralized model predictive control for a class of Interval type-2 discrete-time large-scale systems is considered. The mentioned controller is applied with so-called $H_\infty$ performance to reduce the unknown disturbances, and the adopted modeling is interval type-2 to diminish the uncertainties.

## 4. Numerical example

*Example 1:* In this section, a numerical example shows the applicable of the proposed algorithm for interval type-2 fuzzy T-S large-scale system. The considered fuzzy interval type-2 large-scale system is consist of three subsystems, $S_i, i = 1,2,3$ ($N = 3$), and each subsystem involves two rules, $l = 1,2$. In this example, $N_i = 0.5$, $Q = diag\{1, 1\}$, $M_i = 1$, $\tau_1 = 1$, $\tau_2 = 1.5$, $\tau_3 = 2$. The membership functions are shown by Fig. 2 and Fig. 3. Here, the sampling time is set as Ts = 0.2 minutes. The fuzzy model of the plan is:

$$S_i^l : \begin{cases} \text{IF } z_{i1} \text{ is } \tilde{F}_{i1}^l \text{ and } \ldots \text{ and } z_{ig} \text{ is } \tilde{F}_{ig}^l \\ \text{THEN } x_i(k+1) = A_i^l x_i(k) + B_i^l u_i(k) + E_i^l d_i(k) + \sum_{\substack{j=1 \\ i \neq j}}^{N} g_{ij} x_j(k) \end{cases}$$

So by applying the controller and the membership function we have:

$$x_i(k+1) = \sum_{l=1}^{r_i}\sum_{m=1}^{r_i} w_i^l(z_{iq})h_i^m(z_{iq})\left([\tilde{A}_i^l + \tilde{B}_i^l k_i^m]x_i(k) + E_i^l d_i(k)\right) + \sum_{l=1}^{r_i}\sum_{\substack{j=1\\i\neq j}}^{N} w_i^l(z_{iq})g_{ij}x_j(k)$$

Subsystem $S_1$

$A_{11} = \begin{bmatrix} 0.55 & 0.05 \\ 0 & 0.42 \end{bmatrix}, B_{11} = \begin{bmatrix} 1 \\ 0 \end{bmatrix}, E_{11} = \begin{bmatrix} 0.1 \\ 0 \end{bmatrix}, g_{12} = \begin{bmatrix} 0.08 & 0.05 \\ 0.05 & 0.05 \end{bmatrix},$

$g_{13} = \begin{bmatrix} 0.09 & 0.06 \\ 0.06 & 0.09 \end{bmatrix}, \lambda_1 = 0.5, X_1 = \begin{bmatrix} 0.015 & 0 \\ 0 & 0.015 \end{bmatrix}$

$A_{12} = \begin{bmatrix} 0.4 & 0 \\ 0 & 0.08 \end{bmatrix}, B_{12} = \begin{bmatrix} 0 \\ 1 \end{bmatrix}, E_{12} = \begin{bmatrix} 0 \\ 0.1 \end{bmatrix}.$

Subsystem $S_2$

$A_{21} = \begin{bmatrix} 0.325 & 0 \\ 0.4 & 0 \end{bmatrix}, B_{21} = \begin{bmatrix} 1 \\ -1 \end{bmatrix}, E_{21} = \begin{bmatrix} -0.1 \\ 0 \end{bmatrix}, g_{21} = \begin{bmatrix} 0.1 & 0.1 \\ 0 & 0 \end{bmatrix}$

$g_{23} = \begin{bmatrix} 0 & 0 \\ 0.1 & 0.1 \end{bmatrix}, \lambda_2 = 0.488, X_2 = \begin{bmatrix} 0.018 & 0 \\ 0 & 0.018 \end{bmatrix}$

$A_{22} = \begin{bmatrix} 0.6 & 0.2 \\ 0.1 & 0 \end{bmatrix}, B_{22} = \begin{bmatrix} -1 \\ 1 \end{bmatrix}, E_{22} = \begin{bmatrix} 0 \\ -0.2 \end{bmatrix}.$

Subsystem $S_3$

$A_{31} = \begin{bmatrix} 0.2 & 0.4 \\ 0.2 & 0 \end{bmatrix}, B_{31} = \begin{bmatrix} 1 \\ 1 \end{bmatrix}, E_{31} = \begin{bmatrix} -0.3 \\ 0 \end{bmatrix}, g_{31} = \begin{bmatrix} 0.03 & 0 \\ 0 & 0.02 \end{bmatrix},$

$g_{32} = \begin{bmatrix} 0.1 & 0 \\ 0.1 & 0 \end{bmatrix}, \lambda_3 = 0.487, X_3 = \begin{bmatrix} 0.027 & 0 \\ 0 & 0.027 \end{bmatrix}$

$A_{32} = \begin{bmatrix} 0.3 & 0 \\ 0 & 0.4 \end{bmatrix}, B_{32} = \begin{bmatrix} -2 \\ 1 \end{bmatrix}, E_{32} = \begin{bmatrix} 0 \\ -0.4 \end{bmatrix},$

By considering $\delta(x_i) = \sin(x_i) \in [-1,1]$, the membership functions and parameter uncertainties are:

$\begin{cases} w_i^1(z_{iq}) = 1 - \dfrac{1}{1+e^{x_i+4+\delta(x_i)}} \\ w_i^2(z_{iq}) = 1 - w_i^1(z_{iq}) \end{cases}$,
$\begin{cases} \underline{h}_i^1(z_{iq}) = 1 - \dfrac{1}{1+e^{\frac{-x_i-1.5}{2}}} \\ \overline{h}_i^1(z_{iq}) = 1 - \dfrac{1}{1+e^{\frac{-x_i+1.5}{2}}} \\ \underline{h}_i^2(z_{iq}) = 1 - \overline{h}_i^1(z_{iq}) \\ \overline{h}_i^2(z_{iq}) = 1 - \underline{h}_i^1(z_{iq}) \end{cases}$,
$\begin{cases} \underline{w}_i^1(z_{iq}) = 1 - \dfrac{1}{1+e^{x_i+4-1}} \\ \overline{w}_i^1(z_{iq}) = 1 - \dfrac{1}{1+e^{x_i+4+1}} \\ \underline{w}_i^2(z_{iq}) = \dfrac{1}{1+e^{x_i+4+1}} \\ \overline{w}_i^2(z_{iq}) = \dfrac{1}{1+e^{x_i+4-1}} \end{cases}$

The mentioned parameter uncertainty is assumed as $\delta(x_i) = \sin(x_i) \in [-1,1]$. So we will have:

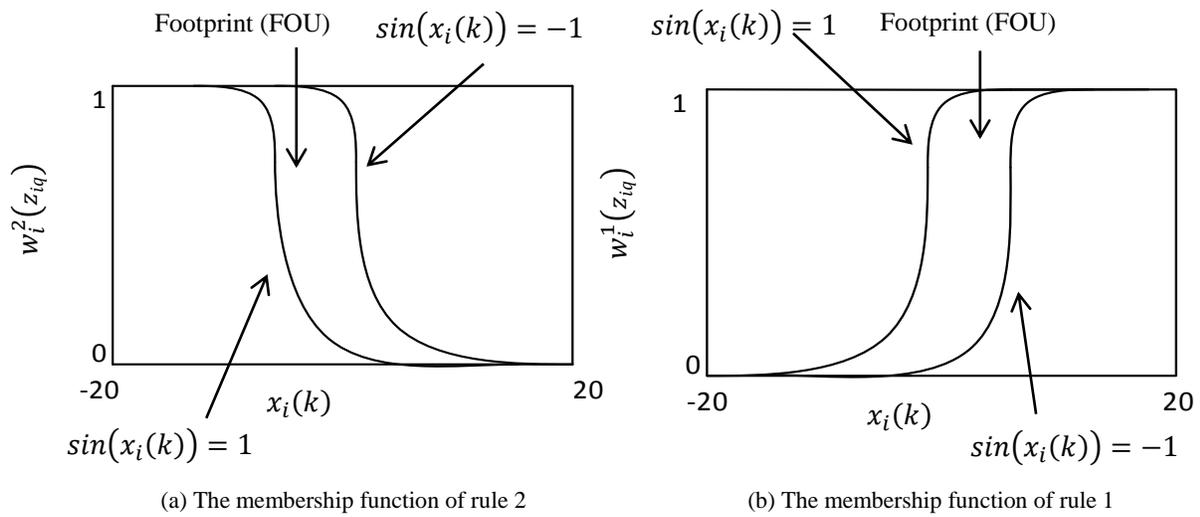

(a) The membership function of rule 2

(b) The membership function of rule 1

Fig. 2 The membership function of IT2 fuzzy model

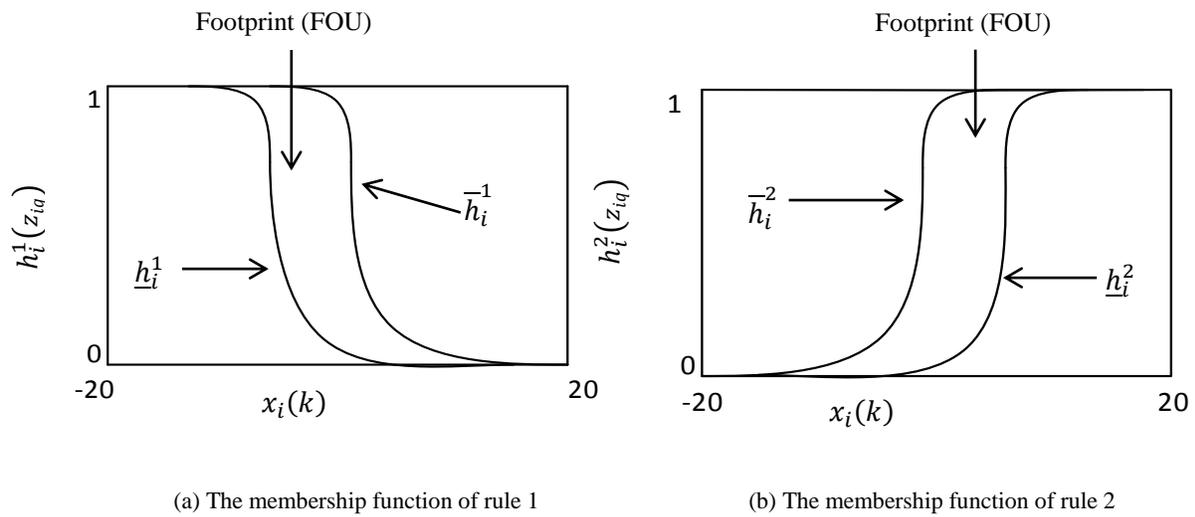

(a) The membership function of rule 1

(b) The membership function of rule 2

Fig. 3 The membership function of IT2 fuzzy controller

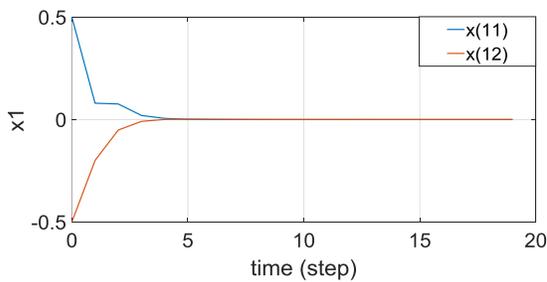

Fig. 4 Trajectories of subsystem 1

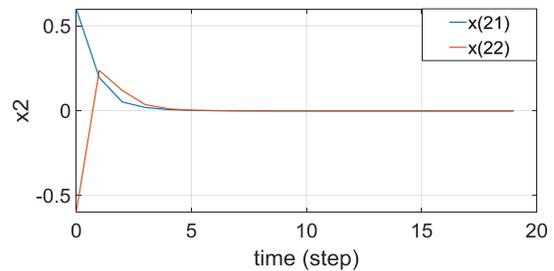

Fig. 5 Trajectories of subsystem 2

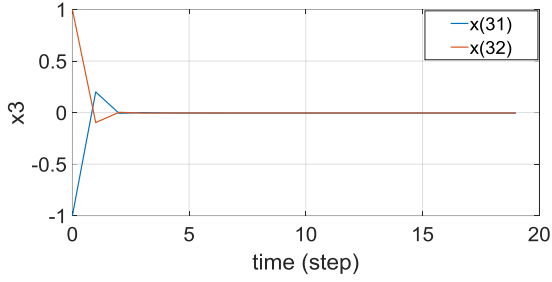

Fig. 6 Trajectories of subsystem 3

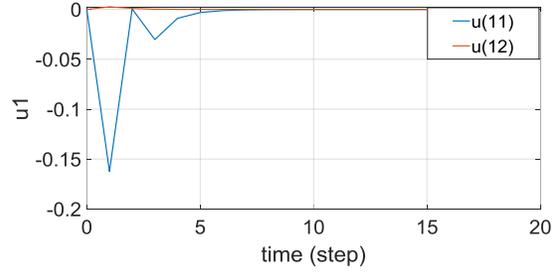

Fig. 7 Trajectories of controller 1

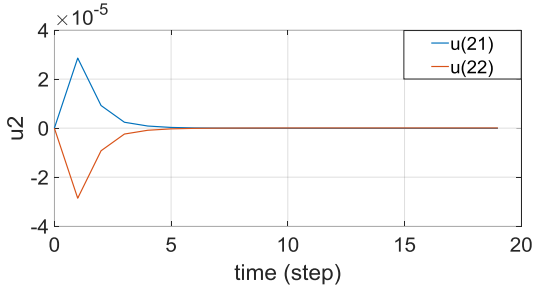

Fig. 8 Trajectories of controller 2

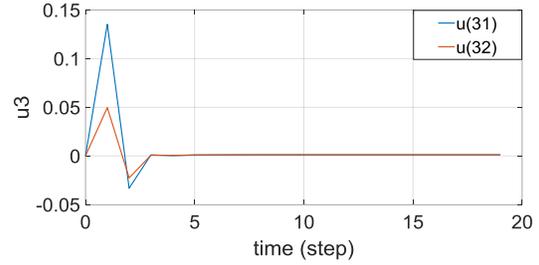

Fig. 9 Trajectories of controller 3

Subsequently, we can obtain the feedback gains:

k11= [-0.549 -0.222]   k12= [-0.0569 -0.799]

k21= [4.794e-05 -4.739e-09]   k22= [1.755e-05 1.138e-05]

k31= [-0.199 -0.111]   k32= [0.073 -0.201]

*Remark 8:* By exemplifying an instance, it has been shown that the proposed algorithm is entirely practical. Here in this example, as is evident in Fig. 4, Fig. 5, and Fig. 6, the trajectories of the three subsystems are leading to zero, and this means that the overall closed-loop system is stable during the time.

*Remark 9:* By noticing to Fig. 7, Fig. 8, and Fig. 9, an important issue is specified. After some whiles, the states of the system have been gotten into a boundary. After that, there is no require to input vector $u_i(k)$, and that means decreasing in costs. This shows the effectiveness of the proposed approach.

*Example 2:* To show the effectiveness of the proposed method, a double inverted pendulum is considered due to the [35]. For convenience, all configurations and parameters are chosen same as [35] and the previous example.

Subsystem $S_1$

$A_{11} = A_{13} = \begin{bmatrix} 1 & 0.005 \\ 0.0262 & 1 \end{bmatrix}, A_{13} = \begin{bmatrix} 1 & 0.005 \\ 0.0441 & 1 \end{bmatrix}, B_{11} = B_{12} = B_{13} = \begin{bmatrix} 1 \\ 0 \end{bmatrix}, E_{11} = E_{12} = E_{13} = \begin{bmatrix} 0.1 \\ 0 \end{bmatrix}, g_{12} = \begin{bmatrix} 0.08 & 0.05 \\ 0.05 & 0.05 \end{bmatrix}, \lambda_1 = 0.5, X_1 = \begin{bmatrix} 0.015 & 0 \\ 0 & 0.015 \end{bmatrix}, H_1 = [1\ 0].$

Subsystem $S_2$

$A_{21} = A_{23} = \begin{bmatrix} 1 & 0.005 \\ 0.0272 & 1 \end{bmatrix}, A_{23} = \begin{bmatrix} 1 & 0.005 \\ 0.0451 & 1 \end{bmatrix}, B_{21} = B_{22} = B_{23} = \begin{bmatrix} 1 \\ 1 \end{bmatrix}, E_{11} = E_{22} = E_{23} = \begin{bmatrix} 0.1 \\ 0 \end{bmatrix}, g_{21} = \begin{bmatrix} 0.08 & 0.05 \\ 0.05 & 0.05 \end{bmatrix}, \lambda_2 = 0.448, X_2 = \begin{bmatrix} 0.018 & 0 \\ 0 & 0.018 \end{bmatrix}, H_2 = [1\ 0].$

$$x_i(k+1) = \sum_{l=1}^{r_i} \sum_{m=1}^{r_i} w_i^l(z_{iq}) h_i^m(z_{iq}) \left( [\tilde{A}_i^l + \tilde{B}_i^l k_i^m] x_i(k) + E_i^l d_i(k) \right) + \sum_{l=1}^{r_i} \sum_{\substack{j=1 \\ i \neq j}}^{N} w_i^l(z_{iq}) g_{ij} x_j(k)$$

$$y_i(k) = H_i x_i(k)$$

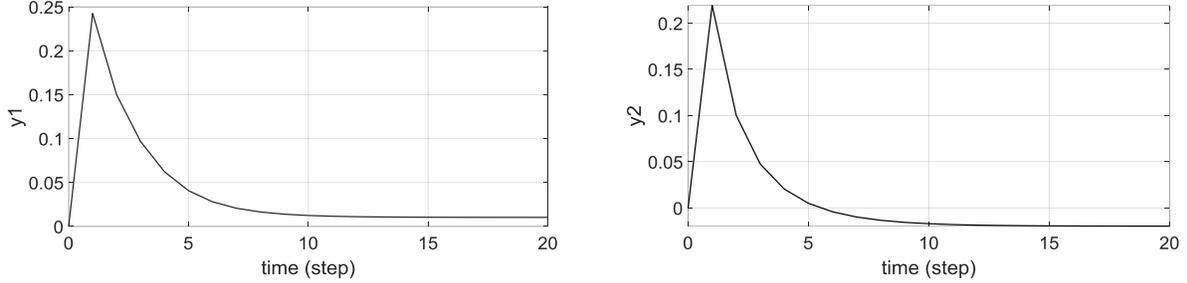

Fig. 10 Output responses of the closed-loop system with the presence of disturbances.

Subsequently, we can obtain the feedback gains:

k11= [-0.071 -0.024]   k12= [-0.0153 -0.219]   k13= [-12.15 -9.585]

k21= [21.794 -41.739]   k22= [-10.75 -21.138]   k23= [-18.255 -32.252]

*Remark 10:* Fig. 10 shows the output responses of the closed-loop discrete-time nonlinear large-scale system with the presence of disturbances. It can be observed that the piecewise controller proposed in this paper based on the fuzzy dynamic model not only stabilizes the original nonlinear large-scale system but also effectively attenuates the disturbances as expected.

*Remark 11:* To evaluate the effectiveness of the proposed controller, a comparison is made due to [36]. In [36], the PI controller is applied to a decentralized large-scale system. In the first set of the experiments, a control scheme using PI controllers, which are currently used in industry, are used. Though this scheme is very simple to implement, its performance is often limited and tuning the P and I gains is a tedious process. In the results of this paper, the proposed controller is used. Experimental results with these control schemes show that the proposed control scheme offers a marked improvement in results.

*Remark 12:* As it is clear due to gain results, to stabilize the mentioned systems in this paper, computed gains have little values. Specifically, in the first example, it is noticed that by gain values that are less than 1, states of the system are stabilized. This means that the proposed method can control the system optimally. In the second example, by comparison with other methods, it is evident that the double inverted pendulum can be stabilized with lower cost and this is the a efficient of the proposed approach.

## 5. Conclusion

In this paper, decentralized robust interval type-2 fuzzy model predictive control for Takagi-Sugeno large-scale systems has been stated. Since almost all real systems have complex and severe nonlinearity, it will be hard, and most of the time, impossible to find the dynamic of systems. Therefore, the interval type-2 fuzzy model of the large-scale system for deducing uncertainties is considered in this paper. The supposed controller for this system is model predictive control, and the type of the controller is decentralized. To reduce the effects of disturbance and uncertainty, the $H_\infty$ performance is involved. Finally, the control gains and other values are calculated through solving LMIs.

**Appendix A.**

*Proof*: By using Schur complement, the inequality (17) is equivalence to:

$$\begin{bmatrix} E_{i\mu}^T X_i E_{i\mu} - \xi_i \lambda_i N_i & \star & \star & \cdots & \star \\ \widetilde{\Theta}_i^T X_i E_{i\mu} & N\sqrt{a}\sum_{\substack{j=1\\i\neq j}}^{N} g_{ij}^T X_j g_{ij} - (-\lambda_i + 1)X_{i\mu} & \star & \cdots & \star \\ g_{ij}^T X_i E_{i\mu} & (1-\sqrt{\alpha})g_{ij}^T X_i \widetilde{\Theta}_i & -(\alpha-1)g_{ij}^T X_i g_{ij} & \cdots & \star \\ \vdots & \vdots & \vdots & \ddots & \vdots \\ g_{iN}^T X_i E_{i\mu} & (1-\sqrt{\alpha})g_{iN}^T X_i \widetilde{\Theta}_i & -(\alpha-1)g_{iN}^T X_i g_{ij} & \cdots & -(\alpha-1)g_{iN}^T X_i g_{iN} \end{bmatrix}$$

$$+ \begin{bmatrix} 0 \\ \widetilde{\Theta}_i^T X_i^T \\ 0 \\ \vdots \\ 0 \end{bmatrix} N X_i^{-T} \begin{bmatrix} 0 & X_i \widetilde{\Theta}_i & \cdots & 0 & 0 \end{bmatrix} \leq 0 \tag{A.1}$$

where $\widetilde{\Theta}_i = \widetilde{A}_{i\mu} + \widetilde{B}_{i\mu} k_{i\mu}$ and if consider $X_i = \xi_i P_i, X_{i\mu} = \xi_i P_{i\mu}$, and $X_j = \xi_i P_j$ and according to the Schur complement, we will have,

$$\begin{bmatrix} \frac{1}{\xi_i} E_{i\mu}^T P_i E_{i\mu} - \frac{\lambda_i}{\eta_i^2} & \star & \star & \cdots & \star \\ \frac{1}{\xi_i} \widetilde{\Theta}_i^T P_i E_{i\mu} & \Xi_i & \star & \cdots & \star \\ \frac{1}{\xi_i} g_{ij}^T P_i E_{i\mu} & \frac{1}{\xi_i}(1-\sqrt{\alpha})g_{ij}^T P_i \widetilde{\Theta}_i & -\frac{1}{\xi_i}(\alpha-1)g_{ij}^T P_i g_{ij} & \cdots & \star \\ \vdots & \vdots & \vdots & \ddots & \vdots \\ \frac{1}{\xi_i} g_{iN}^T P_i E_{i\mu} & \frac{1}{\xi_i}(1-\sqrt{\alpha})g_{iN}^T P_i \widetilde{\Theta}_i & -\frac{1}{\xi_i}(\alpha-1)g_{iN}^T P_i g_{ij} & \cdots & -\frac{1}{\xi_i}(\alpha-1)g_{iN}^T P_i g_{iN} \end{bmatrix} \leq 0 \tag{A.2}$$

where $\Xi_i = \frac{1}{\xi_i} N\widetilde{\Theta}_i^T P_i \widetilde{\Theta}_i + \frac{1}{\xi_i} N\sqrt{a} \sum_{\substack{j=1\\i\neq j}}^{N} g_{ij}^T P_j g_{ij} - \frac{1}{\xi_i}(-\lambda_i + 1)P_{i\mu}$. By multiplying $\begin{bmatrix} d_i^T & x_i^T & x_j^T & \cdots & x_N^T \end{bmatrix}$ and its transpose from both sides of the matrix in the inequality $(A.2)$, respectively:

$$\sum_{i=1}^{N} x_i^T \frac{1}{\xi_i} \left\{ N(\widetilde{A}_{i\mu} + \widetilde{B}_{i\mu} k_{i\mu})^T P_i(\widetilde{A}_{i\mu} + \widetilde{B}_{i\mu} k_{i\mu}) + N\sqrt{a}\sum_{\substack{j=1\\i\neq j}}^{N} g_{ij}^T P_j g_{ij} - P_{i\mu} \right\} x_i$$

$$+ \sum_{i=1}^{N} \left\{ x_i^T \frac{1}{\xi_i}(\widetilde{A}_{i\mu} + \widetilde{B}_{i\mu} k_{i\mu})^T P_i E_{i\mu} d_i + d_i^T \frac{1}{\xi_i} E_{i\mu}^T P_i E_{i\mu} d_i + \frac{1}{\xi_i}\left( \sum_{\substack{j=1\\i\neq j}}^{N} x_j^T g_{ij}^T \right) P_i E_{i\mu} d_i \right.$$

$$+ d_i^T \frac{1}{\xi_i} E_{i\mu}^T P_i(\widetilde{A}_{i\mu} + \widetilde{B}_{i\mu} k_{i\mu})x_i + d_i^T \frac{1}{\xi_i} E_{i\mu}^T P_i \left( \sum_{\substack{j=1\\i\neq j}}^{N} g_{ij} x_j \right) - \frac{\lambda_i}{\eta_i^2} d_i^T d_i + \frac{\lambda_i}{\xi_i} x_i^T P_{i\mu} x_i$$

$$- \frac{1}{\xi_i}(\alpha-1)\left( \sum_{\substack{j=1\\i\neq j}}^{N} x_j^T g_{ij}^T \right) P_i \left( \sum_{\substack{j=1\\i\neq j}}^{N} g_{ij} x_j \right) + x_i^T(1-\sqrt{\alpha})\frac{1}{\xi_i}(\widetilde{A}_{i\mu} + \widetilde{B}_{i\mu} k_{i\mu})^T P_i \left( \sum_{\substack{j=1\\i\neq j}}^{N} g_{ij} x_j \right)$$

$$\left. + (1-\sqrt{\alpha})\frac{1}{\xi_i}\left( \sum_{\substack{j=1\\i\neq j}}^{N} x_j^T g_{ij}^T \right) P_i(\widetilde{A}_{i\mu} + \widetilde{B}_{i\mu} k_{i\mu})x_i \right\} \leq 0 \tag{A.3}$$

by resorting to [37], the inequality $(A.3)$ is:

$$\sum_{i=1}^{N} \frac{1}{\xi_i} \left\{ \left[ (\tilde{A}_{i\mu} + \tilde{B}_{i\mu} k_{i\mu}) x_i + \sqrt{\alpha} \sum_{\substack{j=1 \\ i \neq j}}^{N} g_{ij} x_j \right]^T P_i \left[ (\tilde{A}_{i\mu} + \tilde{B}_{i\mu} k_{i\mu}) x_i + \sqrt{\alpha} \sum_{\substack{j=1 \\ i \neq j}}^{N} g_{ij} x_j \right] - x_i^T P_{i\mu} x_i \right\}$$

$$+ \sum_{i=1}^{N} \left\{ x_i^T \frac{1}{\xi_i} (\tilde{A}_{i\mu} + \tilde{B}_{i\mu} k_{i\mu})^T P_i E_{i\mu} d_i + d_i^T \frac{1}{\xi_i} E_{i\mu}^T P_{il} E_{i\mu} d_i + \frac{1}{\xi_i} \left( \sum_{\substack{j=1 \\ i \neq j}}^{N} x_j^T g_{ij}^T \right) P_i E_{i\mu} d_i \right.$$

$$+ d_i^T \frac{1}{\xi_i} E_{i\mu}^T P_i (\tilde{A}_{i\mu} + \tilde{B}_{i\mu} k_{i\mu}) x_i + d_i^T \frac{1}{\xi_i} E_{i\mu}^T P_i \left( \sum_{\substack{j=1 \\ i \neq j}}^{N} f_{ij} x_j \right) - \frac{\lambda_i}{\eta_i^2} d_i^T d_i + \frac{\lambda_i}{\xi_i} x_i^T P_{i\mu} x_i$$

$$- \frac{1}{\xi_i} (\alpha - 1) \left( \sum_{\substack{j=1 \\ i \neq j}}^{N} x_j^T g_{ij}^T \right) P_i \left( \sum_{\substack{j=1 \\ i \neq j}}^{N} g_{ij} x_j \right) + (1 - \sqrt{\alpha}) \frac{1}{\xi_i} \left( \sum_{\substack{j=1 \\ i \neq j}}^{N} x_j^T g_{ij}^T \right) P_i (\tilde{A}_{i\mu} + \tilde{B}_{i\mu} k_{i\mu}) x_i$$

$$\left. + x_i^T (1 - \sqrt{\alpha}) \frac{1}{\xi_i} (\tilde{A}_{i\mu} + \tilde{B}_{i\mu} k_{i\mu})^T P_i \left( \sum_{\substack{j=1 \\ i \neq j}}^{N} g_{ij} x_j \right) \right\} \leq 0 \qquad (A.4)$$

and the inequality $(A.4)$ is equivalent to:

$$\sum_{i=1}^{N}\left\{\frac{1}{\xi_i}\left[(\tilde{A}_{i\mu}+\tilde{B}_{i\mu}k_{i\mu})x_i+\sum_{\substack{j=1\\i\neq j}}^{N}g_{ij}x_j\right]^T P_i\left[(\tilde{A}_{i\mu}+\tilde{B}_{i\mu}k_{i\mu})x_i+\sum_{\substack{j=1\\i\neq j}}^{N}g_{ij}x_j\right]+x_i^T\frac{1}{\xi_i}(\tilde{A}_{i\mu}+\tilde{B}_{i\mu}k_{i\mu})^T P_i E_{i\mu}d_i\right.$$

$$+d_i^T\frac{1}{\xi_i}E_{i\mu}^T P_i E_{i\mu}d_i+\frac{1}{\xi_i}\left(\sum_{\substack{j=1\\i\neq j}}^{N}x_j^T g_{ij}^T\right)P_i E_{i\mu}d_i+d_i^T\frac{1}{\xi_i}E_{i\mu}^T P_i(\tilde{A}_{i\mu}+\tilde{B}_{i\mu}k_{i\mu})x_i+d_i^T\frac{1}{\xi_i}E_{i\mu}^T P_i d_i$$

$$+d_i^T\frac{1}{\xi_i}E_{i\mu}^T P_i\left(\sum_{\substack{j=1\\i\neq j}}^{N}g_{ij}x_j\right)-\frac{\lambda_i}{\eta_i^2}d_i^T d_i+\frac{1}{\xi_i}(-1+\lambda_i)x_i^T P_{i\mu}x_i$$

$$+\frac{1}{\xi_i}(\alpha-1)\left(\sum_{\substack{j=1\\i\neq j}}^{N}x_j^T g_{ij}^T\right)P_i\left(\sum_{\substack{j=1\\i\neq j}}^{N}g_{ij}x_j\right)-\frac{1}{\xi_i}(\alpha-1)\left(\sum_{\substack{j=1\\i\neq j}}^{N}x_j^T g_{ij}^T\right)P_i\left(\sum_{\substack{j=1\\i\neq j}}^{N}g_{ij}x_j\right)$$

$$+x_i^T\sqrt{\alpha}\frac{1}{\xi_i}(\tilde{A}_{i\mu}+\tilde{B}_{i\mu}k_{i\mu})^T P_i\left(\sum_{\substack{j=1\\i\neq j}}^{N}g_{ij}x_j\right)-x_i^T\sqrt{\alpha}\frac{1}{\xi_i}(\tilde{A}_{i\mu}+\tilde{B}_{i\mu}k_{i\mu})^T P_i\left(\sum_{\substack{j=1\\i\neq j}}^{N}g_{ij}x_j\right)$$

$$\left.+\sqrt{\alpha}\frac{1}{\xi_i}\left(\sum_{\substack{j=1\\i\neq j}}^{N}x_j^T g_{ij}^T\right)P_i(\tilde{A}_{i\mu}+\tilde{B}_{i\mu}k_{i\mu})x_i-\sqrt{\alpha}\frac{1}{\xi_i}\left(\sum_{\substack{j=1\\i\neq j}}^{N}x_j^T g_{ij}^T\right)P_i(\tilde{A}_{i\mu}+\tilde{B}_{i\mu}k_{i\mu})x_i\right\}\leq 0 \quad (A.5)$$

the inequality $(A.5)$ can be written as,

$$\sum_{i=1}^{N}\left\{\frac{1}{\xi_i}\left[(\tilde{A}_{i\mu}+\tilde{B}_{i\mu}k_{i\mu})x_i+E_{i\mu}d_i+\sum_{\substack{j=1\\i\neq j}}^{N}g_{ij}x_j\right]^T P_{i\mu}^+\left[(\tilde{A}_{i\mu}+\tilde{B}_{i\mu}k_{i\mu})x_i+E_{i\mu}d_i+\sum_{\substack{j=1\\i\neq j}}^{N}g_{ij}x_j\right]-\frac{1}{\xi_i}(1-\lambda_i)x_i^T P_{i\mu}x_i\right.$$

$$\left.-\lambda_i\frac{1}{\eta_i^2}d_i^T d_i\right\}\leq 0 \qquad (A.6)$$

we had before $x_i^+=(\tilde{A}_{i\mu}+\tilde{B}_{i\mu}k_{i\mu})x_i+E_{i\mu}d_i+\sum_{\substack{j=1\\i\neq j}}^{N}f_{ij}x_j$. So, we have:

$$\sum_{i=1}^{N}\left\{\frac{1}{\xi_i}x_i^{+T}P_{i\mu}^+ x_i^+-(1-\lambda_i)\frac{1}{\xi_i}x_i^T P_{i\mu}x_i-\lambda_i\frac{1}{\eta_i^2}d_i^T d_i\right\}\leq 0 \qquad (A.7)$$

thus, the RPI property of the set can be obtained if $(A.7)$ holds. Furthermore, the input constraint can be admitted by $(18)$, and the proof is clarified below:

multiplying $diag\{x_i,I\}$ and its transpose from both sides of $(18)$, respectively:

$$\begin{bmatrix} x_i^T Z_i x_i & x_i^T k_{i\mu}^T \\ k_{i\mu}x_i & 1 \end{bmatrix}\geq 0 \qquad (A.8)$$

by the aim of Schur complement to $(A.8)$, then,

$$x_i^T Z_i x_i - (k_{i\mu} x_i)^T (k_{i\mu} x_i) \geq 0 \quad (A.9)$$

As $u_i = k_{i\mu} x_i$, we have:

$$(u_i)^T (u_i) \leq x_i^T Z_i x_i = H_i = positive\ value \quad (A.10)$$

thus $u_i^T u_i \leq H_i$. The proof is, thereby, completed. ∎

**Appendix B.**

*Proof:* Resorting to the Schur complement, and based on the previous proof, the inequality (23) can be written:

$$\begin{bmatrix} E_{i\mu}^T X_i E_{i\mu} - \xi_i \tau_i & \star & \star & \cdots & \star \\ \widetilde{\Theta}_i^T X_i E_{i\mu} & \phi_i & \star & \cdots & \star \\ g_{ij}^T X_i E_{i\mu} & (1-\sqrt{\alpha}) g_{ij}^T X_i \widetilde{\Theta}_i & -(a-1) g_{ij}^T X_i g_{ij} & \cdots & \star \\ \vdots & \vdots & \vdots & \ddots & \vdots \\ g_{iN}^T X_i E_{i\mu} & (1-\sqrt{\alpha}) g_{iN}^T X_i \widetilde{\Theta}_i & -(a-1) g_{iN}^T X_i g_{ij} & \cdots & -(a-1) g_{iN}^T X_i g_{iN} \end{bmatrix} < 0 \quad (B.1)$$

Consider $X_i = \xi_i P_i$, $X_{i\mu} = \xi_i P_{i\mu}$, $X_j = \xi_i P_j$ and $\phi_i = N \widetilde{\Theta}_i^T X_i \widetilde{\Theta}_i + N\sqrt{\alpha} \sum_{\substack{j=1 \\ i \neq j}}^{N} g_{ij}^T X_j g_{ij} - X_{i\mu} + \xi_i Q + k_{i\mu}^T M_i k_{i\mu}$. The inequality $(B.1)$ is:

$$\begin{bmatrix} E_{i\mu}^T P_i E_{i\mu} - \tau_i & \star & \star & \cdots & \star \\ \widetilde{\Theta}_i^T P_i E_{i\mu} & \psi_i & \star & \cdots & \star \\ g_{iN}^T P_i E_{i\mu} & (1-\sqrt{\alpha}) g_{ij}^T P_i \widetilde{\Theta}_i & -(a-1) g_{ij}^T P_i g_{ij} & \cdots & \star \\ \vdots & \vdots & \vdots & \ddots & \vdots \\ g_{ij}^T P_i E_{i\mu} & (1-\sqrt{\alpha}) g_{iN}^T P_i \widetilde{\Theta}_i & -(a-1) g_{iN}^T P_i g_{ij} & \cdots & -(a-1) g_{iN}^T P_i g_{iN} \end{bmatrix} < 0 \quad (B.2)$$

where, $\psi_i = N \widetilde{\Theta}_i^T P_i \widetilde{\Theta}_i + N\sqrt{\alpha} \sum_{\substack{j=1 \\ i \neq j}}^{N} g_{ij}^T P_j g_{ij} - P_{i\mu} + Q + k_{i\mu}^T R k_{i\mu}$. Multiplying $\begin{bmatrix} d_i^T & x_i^T & x_j^T & \cdots & x_N^T \end{bmatrix}$ and its transpose from both sides of $(B.2)$, respectively, we will have:

$$\sum_{i=1}^{N} \left\{ x_i^T \left( N(\tilde{A}_{i\mu} + \tilde{B}_{i\mu} k_{i\mu})^T P_i (\tilde{A}_{i\mu} + \tilde{B}_{i\mu} k_{i\mu}) + N\sqrt{\alpha} \sum_{\substack{j=1 \\ i \neq j}}^{N} g_{ij}^T P_j g_{ij} - P_{i\mu} + Q \right) x_i + x_i^T (\tilde{A}_{i\mu} + \tilde{B}_{i\mu} k_{i\mu})^T P_i E_{i\mu} d_i \right.$$

$$+ d_i^T E_{i\mu}^T P_i E_{i\mu} d_i + \left( \sum_{\substack{j=1 \\ i \neq j}}^{N} x_j^T g_{ij}^T \right) P_i E_{i\mu} d_i + d_i^T E_{i\mu}^T P_i (\tilde{A}_{i\mu} + \tilde{B}_{i\mu} k_{i\mu}) x_i + d_i^T E_{i\mu}^T P_i \left( \sum_{\substack{j=1 \\ i \neq j}}^{N} g_{ij} x_j \right)$$

$$+ x_i^T k_{i\mu}^T R k_{i\mu} x_i - \tau_i d_i^T d_i - (a-1) \sum_{\substack{j=1 \\ i \neq j}}^{N} x_j^T g_{ij}^T P_i g_{ij} x_j$$

$$\left. + x_i^T (1-\sqrt{\alpha}) (\tilde{A}_{i\mu} + \tilde{B}_{i\mu} k_{i\mu})^T P_i \left( \sum_{\substack{j=1 \\ i \neq j}}^{N} g_{ij} x_j \right) + (1-\sqrt{\alpha}) \left( \sum_{\substack{j=1 \\ i \neq j}}^{N} x_j^T g_{ij}^T \right) P_i (\tilde{A}_{i\mu} + \tilde{B}_{i\mu} k_{i\mu}) x_i \right\}$$

$$< 0 \quad (B.3)$$

resorting to [37], the inequality $(B.3)$ is:

$$\sum_{i=1}^{N}\left\{\left[(\tilde{A}_{i\mu}+\tilde{B}_{i\mu}k_{i\mu})x_i + E_{i\mu}d_i + \sum_{\substack{j=1\\i\neq j}}^{N}g_{ij}x_j\right]^T P_{i\mu}^+\left[(\tilde{A}_{i\mu}+\tilde{B}_{i\mu}k_{i\mu})x_i + E_{i\mu}d_i + \sum_{\substack{j=1\\i\neq j}}^{N}g_{ij}x_j\right] - x_i^T P_{i\mu} x_i + x_i^T Q x_i\right.$$

$$+ x_i^T k_{i\mu}^T R k_{i\mu} x_i - \tau_i d_i^T - (a-1)\sum_{\substack{j=1\\i\neq j}}^{N} x_j^T g_{ij}^T P_i g_{ij} x_j + (a-1)\sum_{\substack{j=1\\i\neq j}}^{N} x_j^T g_{ij}^T P_i g_{ij} x_j$$

$$+ x_i^T \sqrt{\alpha}(\tilde{A}_{i\mu}+\tilde{B}_{i\mu}k_{i\mu})^T P_i \left(\sum_{\substack{j=1\\i\neq j}}^{N} g_{ij} x_j\right) - x_i^T \sqrt{\alpha}(\tilde{A}_{i\mu}+\tilde{B}_{i\mu}k_{i\mu})^T P_i \left(\sum_{\substack{j=1\\i\neq j}}^{N} g_{ij} x_j\right)$$

$$\left. + \sqrt{\alpha}\left(\sum_{\substack{j=1\\i\neq j}}^{N} x_j^T g_{ij}^T\right) P_i(\tilde{A}_{i\mu}+\tilde{B}_{i\mu}k_{i\mu})x_i - \sqrt{\alpha}\left(\sum_{\substack{j=1\\i\neq j}}^{N} x_j^T g_{ij}^T\right) P_i(\tilde{A}_{i\mu}+\tilde{B}_{i\mu}k_{i\mu})x_i\right\} < 0 \quad (B.4)$$

now, the inequality $(B.4)$ is:

$$\sum_{i=1}^{N}\left\{\left[\tilde{\Theta}_i x_i + E_{i\mu}d_i + \sum_{\substack{j=1\\i\neq j}}^{N} g_{ij}x_j\right]^T P_{i\mu}^+ \left[\tilde{\Theta}_i x_i + E_{i\mu}d_i + \sum_{\substack{j=1\\i\neq j}}^{N} g_{ij}x_j\right] - x_i^T P_{i\mu} x_i + x_i^T Q x_i + u_i^T R u_i - \tau_i d_i^T d_i\right\} < 0 \quad (B.5)$$

where $\sum_{i=1}^{N} V(x_i^+) - V(x_i) < -\sum_{i=1}^{N} \psi_i(\cdot)$. Thereby, the Proof is completed. ∎

**R**eference